\input psfig
\documentstyle[11pt]{article}
\def\be{\begin{equation}}
\def\ee{\end{equation}}
\def\la{\label}
\def\bea{\begin{eqnarray}}
\def\eea{\end{eqnarray}}
\def\non{\nonumber}
\def\ci{\cite}
\def\la{\label}
\def\fr{\frac}
\def\pp{\partial}
\def\raw{\rightarrow}

\def\lesssim{{_ <\atop{^\sim}}}
\def\bib{\bibitem}
\def\le{\left}
\def\ri{\right}

\def\gm{\gamma}
\def\al{\alpha}

\def\lm{\lambda}

\begin{document}

\begin{flushright}
hep-ph/9910330 \\
IFUNAM-FT-9907 \\

\end{flushright}
\vspace{15mm}

\begin{center}
{\Large \bf Can Moduli Fields parametrize the Cosmological Constant?
} \\
\end{center}
\vspace*{0.7cm}

\begin{center}

{\bf A. de la Macorra\footnote{e-mail: macorra@fenix.ifisicacu.unam.mx} }
\end{center}

\vspace*{0.1cm}
\begin{center}
\begin{tabular}{c}
{\small $$Instituto de F\'{\i}isica, UNAM}\\
{\small Apdo. Postal 20-364}\\
{\small 01000  M\'exico D.F., M\'exico}\\
 \end{tabular}
\end{center}

\vspace{1 cm}

\begin{center}
{\bf ABSTRACT}
\end{center}

{\small  We study the cosmological evolution of string/M moduli fields $T$. We use   T-duality to fix the potential and show that the superpotential $W$ is a function of the duality invariant function $j(T)$ only. If $W$ is given as a finite polynomial of $j$ then
  the moduli fields {\it do not} give an accelerating universe, i.e. they {\it cannot} be used as quintessence.  Furthermore, at $T\gg 1$ the potential is given by a double exponential potential $V\simeq e^{-a\,e^{\sqrt{2}\, T}}$ leading to a  fast decaying behaviour at large times. For moduli potentials with a finite v.e.v. of $T$ the energy density redshift is model dependent but
if $T$ has a finite mass, $m < \infty$, then the  moduli energy density redshifts faster or equal to matter. Only if the moduli mass is infinite can 
the moduli energy density dominate  the universe independently of the initial conditions. 
}

\noindent

\rule[.1in]{14.5cm}{.002in}

\thispagestyle{empty}

\setcounter{page}{0}
\vfill\eject

Can string/M moduli fields  parametrize a slow rolling cosmological constant? Models with a cosmological constant term, intended as a constant vacuum contribution or as a slowly decaying scalar field, have recently received considerable attention for several reasons, both theoretical and observational.

On the observational side, we find direct evidence in recent works on spectral and photometric observations on type Ia supernova \ci{Riess} that a favour positively accelerating universe.  Statistical fits to
several independent astrophysical constraints support these results \ci{Efst}. See however \ci{Blinn} for a different explanation to these observations. More indirect  evidence comes from the observational support for a low matter density universe from X--ray mass estimations in clusters \ci{White}, \ci{Steig}.

 In these works, if the nucleosynthesis limits on the baryonic mass are to be respected, the total matter that clusters gravitationally is
limited to $\lesssim 0.3$. In such a case, a cosmological term would reconcile the low dynamical estimates of the mean mass density with total critical density suggested by inflation and the flatness problem.

The requirements of structure formation models also suggest a cosmological constant term. Simulations of structure formation profit from the presence of matter that resists gravitational collapse and ${\rm \Lambda}$ CDM models provide a better fit to the observed power spectrum of galaxy clustering than does the standard CDM model  \ci{EMS}, \ci{klypin}.

From the theoretical point of view, we have to face the possible conflict between the age of the universe in the standard Einstein--de Sitter model and the age of the oldest stars, in globular clusters.

Estimates of the Hubble expansion parameter from a variety of methods seem to point to $H_0 \approx 70 \pm 10$ km/s/Mpc (a recent review can be found in  \ci{Freed}; see e.g. \ci{Ferr} for specific projects), leading to an expansion age of $t_U \approx 9 \pm 1$ Gyr for a spatially flat universe with null cosmological constant. On another hand, the age of globular clusters have been estimated in the range $\simeq 13-15$ Gyr \ci{Chab0};
although revised determinations based on the Hypparcos distance scale are lower by approximately 2 Gyr \ci{Chab}.

Many models with a scalar field playing the role of a decaying cosmological constant "quintessence" have been proposed up to now. Some of them are specific models motivated by physical considerations but most of them are phenomenological proposals for the desired energy density redshift \ci{RatP}-\ci{cc+redsh}. 

As a first step in the study of these models, the age of the Universe is calculated for several redshift laws of the energy density that resides in the dynamical scalar field \ci{gral}, \ci{Olson}, \ci{PeebR}. Observational consequences of an evolving $\Lambda$ component decaying to matter and/or radiation have been studied in \ci{Freese}, \ci{Wet}, obtaining severe constraints on such models. Quintessence \ci{Cald}  effects on the cosmic microwave background anisotropy are analysed in \ci{Cald} and \ci{Silv} and phenomenological difficulties of quintessence have been studied in \ci{quint}. Constraints on the equation of state of a quintessence--like component have been placed from observational data  \ci{Turn}. Recently, a potential for a cosmologically successful decaying $\Lambda$ term has been constructed in \ci{Albr}. 

 Regarding string/M theory, we have not been able to determine in a unique way the vacuum structure of string/M theory \ci{M}, since there are many consistent and physically inequivalent solutions. However, recent progress has been made in understanding and connecting many of these vacua using duality symmetry and it is hoped that a better understanding of string/M  theory   will yield to a unique low energy limit. 

Even though there are a great number of inequivalent vacua, all these solutions share some general properties and the existence of moduli fields is one of them. Our interest in studying these fields is twofold since the phenomenological consequences of these fields could be relevant for cosmology and on the other hand they could  restrict the possible string vacua and in an extreme case rule out string theory as a physical theory.

The cosmological evolution of scalar fields  with exponential potentials  has been extensively studied in the literature \ci{Wet},\ci{Vexp},\ci{liddle}. These scalar potentials have been motivated by the compactification of the extra dimensions and have been used to describe string/M moduli  \ci{moduli}-\ci{moduli/m2}.  Using $T$-duality invariance, a valid symmetry in string theory  \ci{Tdual}
and M theory \ci{moduli/m1}, we will show that  the string moduli potential   cannot be approximated by an exponential potential if the superpotential has a finite expansion on  the modular invariant function $j(T)$ (see below). The resulting cosmological evolution of string moduli will then greatly differ from an exponential potential one.

To begin,   we will present the Einstein equations for a flat universe with a Friedmann-Robertson-Walker metric in the presence of a barotropic fluid (matter or radiation)  and a scalar filed (i.e. the moduli field) with arbitrary scalar potential. The cosmological evolution of the moduli field $T$ is  determined by the asymptotic behaviour of  the quantity $\lm=-V_{T}/V,\; V_T\equiv \pp V/\pp T$. We will then proceed to rederive and study the non--perturbative scalar potential for the moduli field $T$ and  we will obtain the attractor solutions.   Finally, we discuss the influence of this decaying scalar field on the universe behaviour and we   present our conclusions.

Our starting point is a universe field with a barotropic  energy density, which can be either matter or radiation, and the energy density of the moduli field. The barotropic fluid is described by an energy density $\rho_{\gm}$ and a pression $p_{\gm}$ with a  standard equation of state  $p_{\gm}=(\gm_{\gm}-1)\rho_{\gm}$, where $\gm_{\gm}=1$ for matter and $\gm_{\gm}=4/3$ for radiation. We do not make any hypothesis on which energy density dominates, that of the barotrpoic fluid or that of the moduli field. For a canonically normalized scalar field $\phi$  with a selfinteraction given in terms of the scalar potential $V(\phi)$  but   with gravitational interaction with all other fields (as moduli fields) the equations to be solved, in  a spatially flat Friedmann--Robertson--Walker (FRW) Universe, are given by
\bea
{\dot H}&=&-\frac{1}{2}(\rho_{\gm}+p_{\gm}+{\dot \phi}^2)\non \\
{\dot \rho}&=& - 3 H (\rho + p) 
\la{cosmo} \\
{\ddot \phi}&=& - 3 H \dot\phi  - \frac{dV(\phi)}{d\phi}, \non
\eea
where $H$ is the Hubble parameter, $V(\phi)$ is the scalar field potential and we have taken $8 \pi G = 1$. It is useful to make a change of variables \ci{liddle} $x \equiv  {\dot \phi / \sqrt 6 H}$, $y \equiv  {\sqrt V / \sqrt 3 H}$ and eqs.(\ref{cosmo}) become
\bea
x_N&=& -3 x + \sqrt {3 \over 2} \lambda\,  y^2 + {3 \over 2} x [2x^2 + \gm_\gm (1 - x^2 - y^2)]  \non \\
y_N&=& - \sqrt {3 \over 2} \lambda \, x\, y + {3 \over 2} y [2x^2 + \gm_\gm (1 - x^2 - y^2)]
 \la{cosmo1} \\
H_N&=& -{3 \over 2} H [\gm_\gm (1-x^2-y^2) + 2x^2] \non
\eea
where $N$ is the logarithm of the scale factor $a$, $N \equiv ln (a)$, $f_N\equiv df/dN$ for $f=x,y,H$ and $\lambda (N) \equiv - V_{\phi} / V$. Notice that all model dependence in eqs.(\ref{cosmo1}) is through the quantities $\lm (N)$ and the constant parameter $\gm_{\gm}$. Eqs.(\ref{cosmo1}) must be supplemented by the Friedmann or constraint equation for a flat universe ${\rho_{\gm} \over 3 H^2} + x^2 + y^2 = 1 $ and they are valid for any scalar potential as long as the interaction between the scalar field and matter or radiation is  gravitational only. This means that it is possible to separate the  energy  and pression densities  into contributions from each component, i.e. $\rho=\rho_{\gm}+\rho_{\phi}$ and $p=p_{\gm}+p_{\phi}$, where   $\rho_{\phi}$ ($p_{\phi}$) is the energy density (pression) of the scalar field. We do not assume any equation of state for the scalar field. This is indeed necessary since one cannot fix the equation of state and the potential independently. For arbitrary potentials the equation of state for the scalar field  $p_{\phi}=( \gm_{\phi} -1)\rho_{\phi}$ is determined once $\rho_\phi, p_\phi$ have been obtained. Alternatively we can solve for $x, y$  using eqs.(\ref{cosmo1}) and  the quantity $\gm_{\phi}=(\rho_{\phi}+p_{\phi})/\rho_{\phi}=2x^2/(x^2+y^2)$ is, in general, time or scale dependent.

A complete analysis of the solutions of eqs.(\ref{cosmo1}) has been  obtained \ci{mioscalar} and the values of $x, y$ at late times depend on the asymptotic behaviour of $\lm$. If $\lm \raw 0$ then $x \raw 0$,  $y\raw 1$ and $H\raw cte$ becoming the scalar field a "true" cosmological constant. If $\lm \raw \infty$ then depending on whether $\lm$ oscillates or not we will have different asymptotic behaviour. If the v.e.v. is at $\phi \raw \infty$ then $\lm$ will not oscillate and $x\propto y \propto \lm^{-1} \raw 0$ and $\Omega_\phi \raw 0$. If, on the other hand,  the v.e.v. of $\phi$ is finite, $\lm$ will oscillate and approach infinity. In this class of potentials we can expand around the minimum with a leading term $V(\phi)=V_0 \, (\phi-\phi_0)^n$ giving an equation of state  with $\gm_\phi = 2n/(2+n)$ \ci{scherr}-\ci{mioscalar}.   If $\gm_\phi < \gm_\gm$ ( $\gm_\phi > \gm_\gm$) then  $\Omega_{\phi} \raw 1$ ($\Omega_\phi \raw 0$) and for $\gm_\phi = \gm_\gm$,   $ \Omega_{\phi} $   goes to a finite ($\neq 0,1$) constant value. We have therefore seen that asymptotic limit of  the quantity $\lm$ determines the cosmological evolution of the scalar field and the redshift of the energy density.

Let us now rederive the non--perturbative potential for string moduli. These fields arise in string theory when the extra dimensions are compactified and they parametrize the  geometry or complex structure of the compactified manifold.   For string theory compactified on an orbifold there are at least 3 moduli $T$  that represent the radii  of the manifold ($T = R^2 + i B$ with $R$ the radius of compactification and $B$ an axion field).  Moduli fields  do not have any gauge interaction, they interact only gravitationally with all other fields and as long as supersymmetry is preserved they have a completely flat potentials, i.e. $V(T)\equiv 0$. However, once supersymmetry is broken they acquire a steep potential that can be determined using T-duality invariance. T-duality  has been proven to be present in  low energy string theories \ci{Tdual} aaaand in M theory \ci{moduli/m2}. This symmetry ensures that the contribution of the Kaluza-Klein modes, which appear because the fundamental object is a string and not a point particle, are taken into account in the 4-D theory. The $T$-duality symmetry is given by an $SL(2,Z)$ group where the modulus field $T$ transforms as $T \rightarrow T' = (a T - i b) / (i c T + d)$ with $ad-bc=1$ and $a,b,c,d \ \epsilon \ Z$ \ci{sl2z}. This symmetry   gives the famous $T\rightarrow 1/T$ (i.e. $R \rightarrow 1/R$) invariance and it has two duality invariant points $T=1$ and $T=\delta\equiv e^{i\pi/6}$ plus $SL(2,Z)$ related points. Due to duality invariance we will only need to consider the parameter space with $T >1$.

Another important symmetry of the  4-D string theory is supersymmetry. Global supersymmetry is given in terms of two functions, the Kahler potential $K(T,\bar T)$, an analytic real function, and  the superpotential $W(T)$  an holomorphic function that depends only on $T$ (not on $\bar T$).  The Kahler potential gives the kinetic term
\be
{\cal L}_k = K_T^T (\partial_\mu T) (\partial^\mu \bar T)
\la{Kin}
\ee
with $ K_T^T \equiv \frac{\partial^2 K}{\partial T \partial \bar T}$ and the scalar potential is 
\be
V = (K^{-1})_T^T \vert W_T \vert^2
\la{V}
\ee 
with $ W_T \equiv\frac{\partial W}{\partial T}$. In order to have a duality invariant model $K$ and $W$ must be duality invariant in global susy. For $K, W$ duality invariant, the derivatives of $K$ and $W$ are not duality invariant but the scalar potential  as given in eq.(\ref{V}) is  indeed invariant. For local susy it is only  the combination $G\equiv K+log|W|^2$ that must be duality invariant and the scalar potential is given in terms of the covariant derivative  $G_T=\fr{\pp G}{\pp T}=K_T+\fr{W_T}{W}$ by $V=e^G (G_T(K^{-1})_T^T G^T-3)$. Notice that duality invariance is   more  restrictive for global susy   than for local susy since in the former  case $K,\,W$ must be independently duality invariant.

In string theory, the kinetic term has been calculated  in the large $T$ limit and it is given by a non linear sigma model 
\be
{\cal L}_k = \frac{1}{T_r^2} (\partial_\mu T) (\partial^\mu \bar T)
\la{kin}\ee 
where $T_r \equiv T + \bar T$ and from eq.(\ref{Kin})  we have $K_T^T= 1/T_r^2$.  The Kahler potential must then be $K=-log(T + \bar T) + log(f(T)) + log(\bar f(\bar T))$. In order to have $K$ duality invariant the function $f(T)$ must transform  as modular function with weight $m=1$ (under a duality transformation  a modular function transforms as $f \raw (ic+d)^m f$ with $m$ the modular weight) to cancel the transformation of $T_r$. If we do not wish to introduce any spurious zeros or poles in the fundamental domain then $f=\eta(T)^2$, where $\eta(T)$ is the  Dedekind eta function defined by $\eta(T) \equiv q^{1/24} \prod_n (1 - q^n),\, \ q  \equiv e^{-2 \pi T}$ with  modular weight $m=1/2$.

The superpotential must be a duality invariant function and it is therefore  given in terms of the absolute invariant function $j(T)$, since all invariant functions can be expressed as a rational function of $j(T)$ (see appendix for definitions of modular functions and useful relations between them but for completeness see \ci{modfunc} ). All string vacua will then have a Kahler potential and superpotential of the form
\be
K(T,\bar T)=-log(T_r|\eta(T)|^4),  \hspace{2cm} W(T)=W(j(T))
\ee
respectively. Notice that $W$ depends on $T$ only through $j(T)$. This choice of $K$ and $W$ is valid for global and local susy although it is more common to write in local susy $K_L=-log(T_r)$ and a superpotential $W_L=W\, \eta^{-2}$ with $G=K_L+log|W_L|^2=K+log|W|^2$. In this later case neither $K_L$ nor $W_L$ are duality invariant but $G$ is indeed invariant.

At large $T$ one has $j(T)\simeq e^{2\pi T}$ and for $W=j(T)^{-a}$ one  obtains a superpotential $W\simeq e^{-2a\pi T}$ and a scalar potential
\be
V\simeq T_r^2 \,e^{-2\pi a T_r}
\la{vexp}
\ee
leading to the well known exponential potential for moduli fields. Exponential potentials have been extensively studied in cosmology and lead to a nonvanishing contribution of the energy density at late times  and to other interesting properties \ci{Wet},\ci{Vexp},\ci{liddle}. However,  we would like to point out that for a finite expansion of $W$ on $j$,  string moduli fields  {\it do not} have exponential potentials because their kinetic term is non-canonically normalized (c.f. eq.(\ref{kin})). In fact,  taking $T$ real, the canonically normalized field $\phi=log(T)/\sqrt{2}$ the potential in eq.(\ref{vexp}) is $V(\phi)\simeq e^{2\sqrt{2}\,\phi} \, e^{-2\pi a\, e^{\sqrt{2}\, \phi}}$, i.e. it is a double exponential potential.

 Different string vacua will have different superpotentials $W(j(T))$, however, we can study them in general using the properties of the invariant function $j(T)$ and the physical condition that the potential $V$ should vanish at its minimum. This condition is necessary if we want the moduli potential to represent a slow varying cosmological constant. If the potential does not vanish  at its minimum then we would need to explain the smallness of the vacuum energy and we would introduce a fine tuning problem. We therefore demand the scalar potential and its first derivative to vanish at the minimum, i.e. $V|_{min}=V'|_{min}=0$.  These conditions  require  $W'=0$ as seen from eq.({\ref{V}), where from now on the prime denotes derivative with respect to $T$.  What are the zeros of $W'$? We will distinguish two different kinds of zeros: at finite $T=T_0$ and at $T=\infty$.    The function $W'$ will always vanish at the dual invariant points $T=1,\,\delta=e^{i\pi/6}$, as  any modular invariant function, but it may also vanish  at a model dependent point $T_0$.

Let us now proceed to study the cosmological evolution of the moduli field. We need to solve eqs.(\ref{cosmo1}). Since all model dependence in eqs.(\ref{cosmo1}) is given through $\lm=-V'/V$   its asymptotic behaviour will  determine the cosmological evolution of the moduli field at late times \ci{mioscalar}. We will show that in all cases $\lm$  will tend to infinity  for $T$  canonically normalized if the superpotential $W$ is given as a finite expansion on $j$. Here we will mainly consider this kind of superpotentials. For finite values of $T$, the field will oscillate around its v.e.v. and $\lm$ will also oscillate while approaching infinity. In this class of models the redshift of energy density of the moduli field is model dependent \ci{scherr},\ci{mioscalar}. On the other hand, for $T\raw \infty $ then $\lm\raw \infty$ without oscillating and the energy density of the moduli field will redshift  faster than the barotropic fluid for all models. For the potential given in (\ref{V}), we have
\be
\lm=-\fr{V'}{V} =-\le(\fr{1}{T_r}+\fr{W''}{W'}\ri)= -\le(\fr{1}{T_r}+\fr{j''}{j'}+j' \fr{W_{jj}}{W_j} \ri)
\la{lm}
\ee
where we have used that $W$ is a function of $j(T)$ only and $W_j\equiv dW/dj$. The absolute invariant function $j(T)$ has its only zero (triple zero) at $T=\delta$ plus $SL(2,Z)$ related points  and it has a  pole at the cusp point $T=\infty$ and by duality at $T=0$. The  derivative of $j$, $j(T)'$, vanishes only at the dual invariant points $T=1,\delta$ and it has a pole at $T=\infty$ and $ T=0$.

In order to determine the asymptotic value of $\lm$ at late times we will study the limiting behaviour of $j''/j'$ and $j'W_{jj}/W_j$. From the functional dependence of $j$ on $T$ we obtain the following limits  (see appendix). For  $T\raw 1$ we have $j''/j' \raw -\infty$,  for $T\raw \delta$ we find $\j''/j' \raw \infty$  and for a model dependent point $T=T_0 (\neq 1,\delta)$ then the value of $j''/j'$ is  model dependent but will be, in general, different than zero or infinity.  Finally, for $T\raw \infty$ we have a constant value, $j''/j' \raw 2\pi$. 

Now, we consider  the last quantity in eq.(\ref{lm}). Without loss of generality if $<T>$ is finite, we can expand $W$ around the  zeros of $W'$ that we wish to analyse, i.e. $W=\Sigma_n a_n (j(T)-j_0)^n$ with  $j(T_0)\equiv j_0$ and $V(T_0)= W'(T_0)=0$.  Since in the asymptotic limit $T\raw T_0$ and $j(T) \raw j_0$ the leading term of $W_j$ and $W_{jj}$ is the one with the smallest exponent, say  $n_0$, giving $W_{jj}/W_j\simeq a_0(n_0-1)/(j-j_0)$. For $T_0 \neq 1, \delta$ but finite  we have $j'\neq 0$, and  $j'W_{jj}/W_{j}\raw  \infty $ in the limit $T\raw T_0$. If the minimum is at $T=\delta$, then $j(T_0)=j_0=0$ and since $j$ has a triple zero while $j'$ only a double zero $j'W_{jj}/W_j\raw a_0(n_0-1)j''/j'$ also approaches infinity while for $T_0=1$ we have two possibilities. Since $W'(T=1)=j'(T=1)=0$ for any value of $j_0$ we could take $j_0=0$ and then $j'W'_{jj}/W_j\raw 0$. On the other hand, we could take $j_0=j(1)=1728$ in which case $j'W_{jj}/W_j\raw  \infty $. Finally,  for $T\raw \infty $ we find $j'W_{jj}/W_{j}=a_0(n_0-1)2\pi$ which is the only case where it remains finite.

 We have found, so far,  that  $\lm \rightarrow \infty$ for finite values of $T$ (i.e. $T=1,\delta,T_0$) and it tends to a constant $\lm \raw  -2\pi(a_0( n_0-1)-1)$  for  $T\rightarrow \infty$.   However, the moduli field $T$ has non canonically normalized kinetic terms and  eqs.(\ref{cosmo}) or (\ref{cosmo1}) are valid only for canonically normalized fields. The canonically normalized field is $\phi=ln(T)/\sqrt{2}$ (we are taking $T$ to be a real field since its imaginary part is an axion field).  In terms of $\phi$ the relevant cosmological parameter is $\lm_{c.n.}=-(\pp V/\pp \phi) /V=-\sqrt{2}\,T \,\lm $ giving  $\lm_{c.n.} \rightarrow \infty$ in all cases, i.e. for $T \raw 1,\delta,T_0,\infty$. Besides the asymptotic limit  it is also the oscillating behaviour of $\lm$ that is important in determining the cosmological evolution of the moduli field.  In table 1 we show the behaviour of several cosmological relevant quantities: $\lm_{c.n.}, \gm_T,  \Omega_T=\rho_T/\rho$ and $\al$ for the different v.e.v. of $T$.  The acceleration parameter $\al$ is defined as \ci{mioscalar}
\be
\alpha\equiv \frac{\rho +3p}{(3\gm_{\gm}-2)\rho}=\frac{3\gm-2}{3\gm_{\gm}-2}
\la{al}
\ee
with $\gm =(\rho+p)/\rho$. 
If $\alpha=1$ then the acceleration  of the universe is the same as that of the barotropic fluid and any deviation of $\alpha$ from one implies a different cosmological behaviour of the universe due to the contribution of the  scalar field. A positive accelerating universe requires a negative $\alpha$ while for $0<\alpha<1$ the acceleration of the universe is negative (deceleration) but smaller than that of the barotropic fluid. For $\alpha>1$ the deceleration is larger than for the barotropic fluid.  In  terms of the standard deceleration parameter $q\equiv - \frac{\ddot{a} a}{\dot{a}^2}$ one has $\al=\frac{2q}{3\gm_{\gm-}2}$ or in terms of $x, y$  one finds $\alpha=1-\fr{3\gm_{\gm}}{3\gm_{\gm}-2}(y^2-\frac{2-\gm_{\gm}}{\gm_{\gm}}x^2)
=1-3\Omega_T \fr{\gm_{\gm}-\gm_T}{3\gm_{\gm}-2} $.  

Our analysis so far has considered finite expansion of the superpotential only. However, a quite interesting case arises for a superpotential\footnote{A complete analysis will be presented elsewhere \ci{miolog}} $W=Log[j(T)]^a$. In this case, the leading term  in $j$ and $j'$ for  $T \gg 1$ in the scalar potential $V=a^2 T_r^2 |j'/j|^2 |Log(j)|^{2(a-1)}$ cancels out  and $V \sim T^{2a}=e^{a2\sqrt{2}\,\phi}$.  This superpotential is the only one which allows for an exponential potential for the moduli field and it has $\lm{c.n.}=-a 2 \sqrt{2}$. The vanishing  of the potential at infinity requires $a<0$.

When the minimum of the potential is at $T=\infty$, $\lm$ approaches infinity and does not oscillate. In this case the energy density of the moduli tends to zero and $x, y \raw 0$ in a few e-folds. This result excludes string moduli field as a slow varying cosmological constant when the   moduli field tends to infinity. It is common in the literature to use exponential potentials to parameterize  moduli fields in the $T\gg 1$ region, however, we have shown that for finite expansion of $W$ on $j$, the correct asymptotic behaviour of moduli fields is to have a double exponential potential ( $V=V_0 e^{-a e^{\sqrt{2} \phi}}$)  giving a completely different cosmological evolution for the moduli field.  In fig.(1) we show the evolution of different cosmological parameters for a superpotential $W=j(T)^{-1}$ which gives a scalar potential $V= T_r^2 |j'/j^{-2}|^2$. In the region $T \gg 1$ one has $V \simeq T^2 e^{-4\pi\,T}, \,T=e^{\sqrt{2}\,\phi}$. We can see that  the behaviour of $x,y$ is to rapidly approach zero with $\lm$ constant. This is the scaling regime. Afterwards, $\lm$ increases and tends to infinity while $x,y$ after a subsequent  increase end up going to zero with $x \raw \sqrt{\fr{3}{2}}\;\fr{\gm_{\gm}}{\lm},\, y \raw \sqrt{\frac{3(2-\gm_{\gm})\gm_{\gm}}{2\lm^2}}$ \ci{mioscalar}. The behaviour of the moduli field with the complete duality invariant potnetial is very similar to the double exponential case shown in \ci{mioscalar}. The evolution of the Hubble parameter $H$ approaches  $H_m$ (i.e. for matter fields) after a few e-folds while the accelerator parameter $\al$ goes to one.  The acceleration of the universe is then not affected by the moduli field. The same conclusion holds for the expansion rate of the universe given by $\gm=\rho/(\rho+p)=(\al-2)/3$.

\begin{figure}
\psfig{file=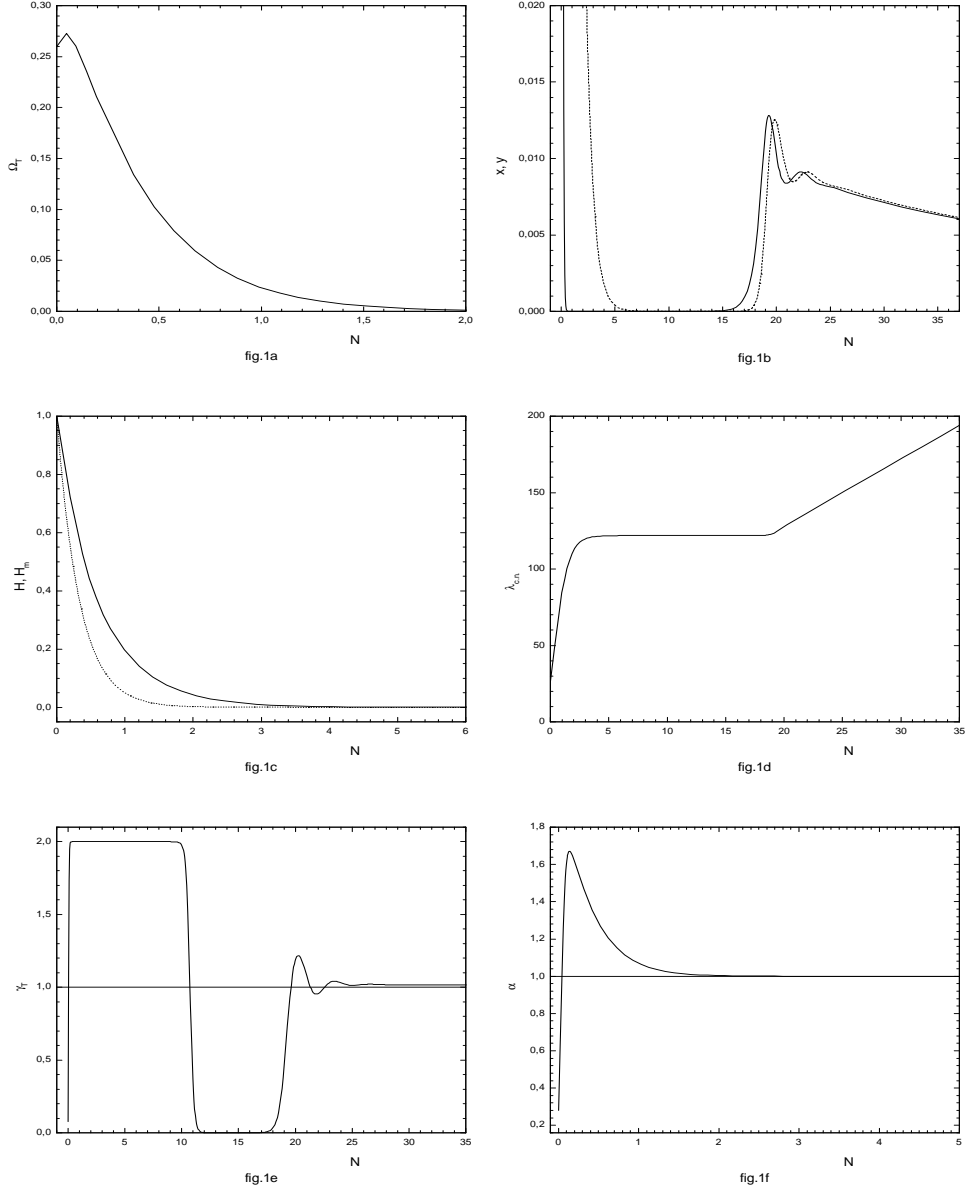,width=16cm,height=20cm}
\caption{
\footnotesize{
We show different cosmological parameters for a potential $V=c\, T_r^2 |j'/j^{-2}|^2$ with $T \raw \infty$ at late times. The initial conditions are $x_0=0.1, y_0=0.5$ and $H_0=1$. In fig.2a we show the evolution of $\Omega_T$. In fig.2b we see the behaviour of x (dot line) and y (solid line). In fig.2c we plot the Hubble parameter (dash-dot line) compared to a matter dominiated universe $H_m$  (dot line). In fig.2d we have $\lm{c.n.}$. Finally we show in fig.2e and 2f the  $\gm_{c.n.}$ $\al$ parameters respectively.}}
\end{figure}

If the   v.e.v. of $T$ is finite  the cosmological evolution of the moduli field is model dependent. We have seen that in all these cases $\lm \raw \infty$ with an oscillating behaviour. This is of course due to the oscillating behaviour of the field around the minimum of the scalar potential. Without loss of generality we can expand the scalar potential around the v.e.v. of the canonically normalized scalar field $\phi$ and keeping only the leading term one has $V \simeq V_0 (\phi-\phi_0)^n$.  This kind of potentials have an average equation of state parameter $\gm_{\phi} \raw 2n/(2+n)$ \ci{scherr},\ci{mioscalar}. The condition to have a finite mass (i.e. $\fr{\pp^2 V}{\pp \phi^2}= m^2 < \infty$)  requires $n \ge 2$ which implies that $\gm_\phi \ge 1$. This means that the redshift of the moduli field will be faster or at best equal to that of  matter fields and  therefore it will not parameterize a slow rolling cosmological constant. However, if we don't impose a finite moduli mass then the condition $V|_{min}=V'|_{min}=0$ only imposes the constraint $n> 1$. For  $1< n < 2$ we have $2/3 < \gm_\phi < 1$ giving a moduli redshift slower than for matter fields. So, even though a slow varying cosmological constant can be obtained and it could help to solve the universe age problem, the condition $n\ge 1$ implies that it will not positively accelerate the universe (i.e. $\al < 0$) as required by the recent supernova results. For $n < 0$ the moduli field will tend to infinity and the correct scalar potential is given by the double exponential potential (see above). In terms of a superpotential $W=(j(T)-j_0)^{n_0}$ and the scalar potential  $V \simeq V_0 (\phi-\phi_0)^n$ the relationship between $n_0$ and $n$ is: $n=2$ for for $T=1$ independently of the value of $n_0$, $n=2(3n_0-1)$ for $T_0=\delta$ and $n=2(n_0-1)$ for a model dependent value $T_0$. The difference in the relationship between $n$ and $n_0$ is due to the different multiplicty of the zero around the value of $T_0$ that we are expanding. If the superpotential has a finite expansion on $j$ then the moduli fields will not have a  scalar potential with negative powers of $T$, i.e. $V \sim T^b, \, b <0$, since the leading term in $j$ and $j'$ is given by $1/q=e^{2\pi\,T}$ and at large $T$ any dependence in $V$ on $j$ or $j'$ will   give the main contribution.

\footnotesize{
\begin{center}
\begin{tabular}{|c|c|c|c|c|}
\hline
 $T$     &       $\lm_{c.n.} = -V_\phi/V$  &   $\gm_T $ & $\al$ & $\Omega_T$ \\
\hline
 1   &         $ \infty  $     & 1 & 1 & $0< \Omega_T < 1$ \\
\hline
 $ \delta=e^{i\pi/6}$    &         $ \infty$   & $2-\fr{2}{3n_0} $  & $1-\Omega_T (\fr{2}{n_0} -3) > 0$ &  $0 \,(1)$ for $n_0 > 2 \;(< 2)$\\
 \hline
 $ 1<  T_0 < \infty$    &        $  \infty$    & $2-\fr{2}{n_0} $    &  $1-3\Omega_T (\fr{2}{n_0}-1) > 0$ & $0 \,(1)$ for $n_0 > 2/3 \;(< 2/3)$ \\
\hline
 $\infty$    &       $ T \sqrt{2}\,2\pi n_0 \raw \infty$ & $\gm_{\gm}$ & 1 & 0
\\ 
\hline
$\infty$ & $-a 2 \sqrt{2}$ & $\gm_{\gm}$ or $8a^2$    &  1 or $8a^2-2$   & $3\gm_{\gm}/8a^2$ or 1
\\
\hline
\end{tabular}
\end{center}
Table 1. We show the  asymptotic behaviour of $\lm_{c.n.},\, \gm_{T},\, \al$ and $ \Omega_{T}$ for different v.e.v. of $T$ and a superpotential with leading term $W=(j(T)-j_0)^{n_0}$ and we have taken $\gm_{\gm}=1$ in $\al$. In the 
last line we have taken a superpotential $W=Log[j(t)]^a,\,a < 0$ and the first quantity in $\gm_T, \al, \Omega_T$ corresponds to $a^2>3\gm_{\gm}/8$ while the second one for $a^2 < 3/4$ respectively.
}}

We give two examples of oscillating moduli fields. In fig.2 we show the evolution of different cosmological parameters for a superpotential $W=j(T)^2$ and $V=c\,  T_r^2 |2j'j|^2$, with\footnote{We took a small $c$ to allow the initial value of $T_i$   not to be  too close to $T_0=1$. For $c=1$ 
one has $T_i=1.007$ while for $c=10^{-13}$ one has $T_i=2.04$.}$c=10^{-13}$ and initial conditions $x_0=0.1, y_0=0.5, H_0=1$. The scalar potential vanishes at the dual invariant points and we study the behaviour around $T_0=1$. Since in this case $V \propto (\phi-1)^2$, i.e. $n=2$, the moduli energy density should redshift as matter in the asymptotic region and as can be seen from fig.2a. We also see in fig.2b that $x, y$ approach zero quite rapidly and they start oscillating when the scaling period is over, as can be seen by comparison with fig.2e. The acceleration parameter tends to one after a few e-folds and $\gm_T$ ends up oscillating with an average  $<\gm_T>=1$.

\begin{figure}
\psfig{file=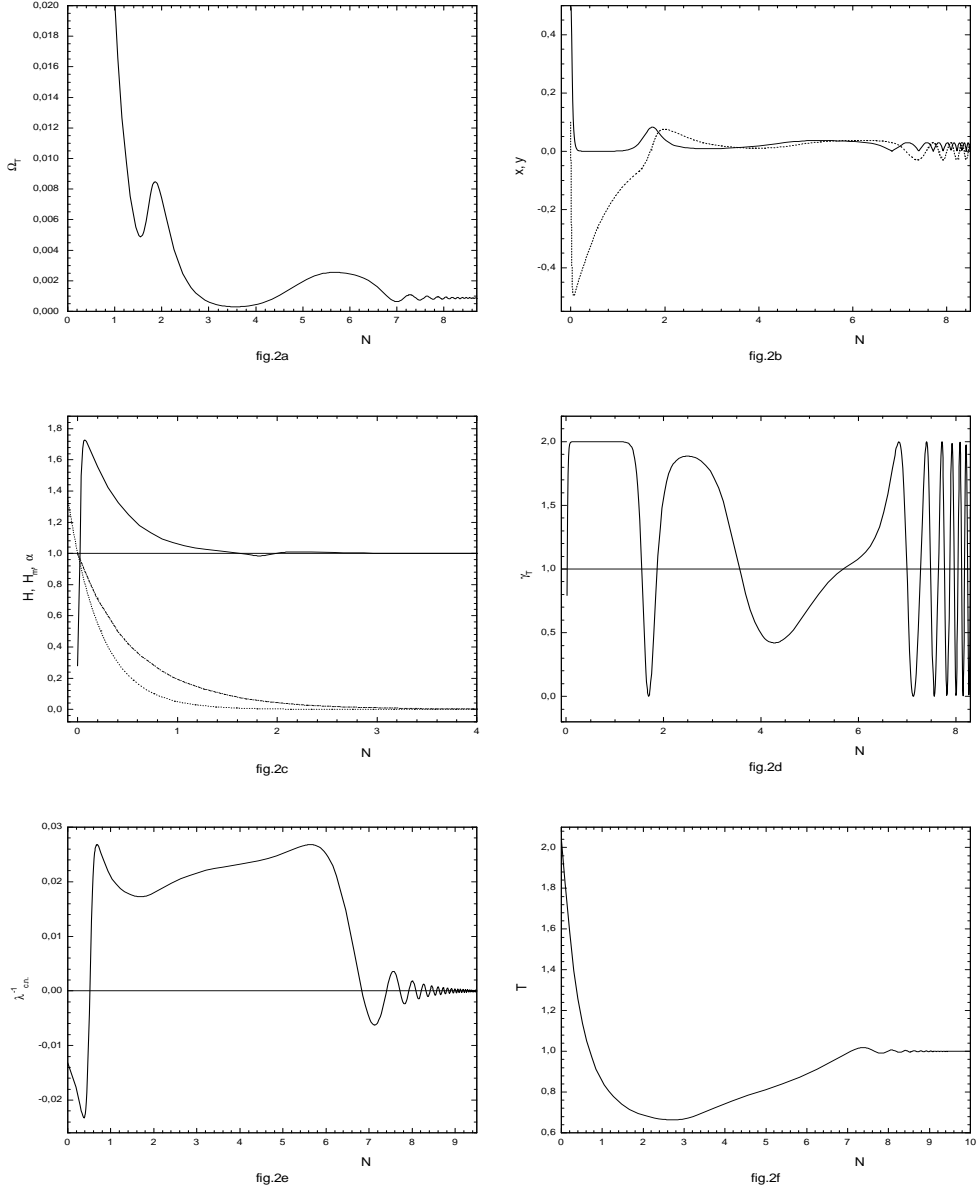,width=16cm,height=20cm}
\caption{
\footnotesize{
We show the evolution of different cosmological parameters for a potential $V=c\, T_r^2 |2j'j|^2$ with $T \raw 1$ at late times. The initial conditions are $x_0=0.1, y_0=0.5$ and $H_0=1$. In fig.1a we show the evolution of $\Omega_T$. In fig.1b we see the behaviour of x (dot line) and y (solid line). In fig.1c we plot the Hubble parameter (dash-dot line) compared to $H_m$ (dot line) and the $\al$ parameter (solid line).In fig.1d we have $\gm_T$. Finally we show in fig.1e and 1f the  $\lm_{c.n.}$ and the evolution of $T$.}}
\end{figure}

As our second example for an  oscillating field we
 show, in fig.3, the cosmological evolution for a superpotential  $W=(j-j_0)^{8/15}$ with  a scalar potential  $V=c\, (8/15)^2  T_r^2 |j'|^2|(j-j_0)|^{-7/15}$, with   $j_0=j(T=2),\, c=10^{-9}$ and initial conditions $x_0=0.1, y_0=0.5, H_0=1$. The scalar potential vanishes at the dual invariant points and at $T_0=2$ and we study the behaviour around $T_0=2$.  Since in this case $V \propto (\phi-1)^{6/5}$, i.e. $n=6/5$, the equation of state parameter is $<\gm_T>=3/4$ and  the moduli energy density  must redshift slower than matter. This can  be seen in fig.3a.  Notice in fig.3e and fig.3f that the average $\gm_T$ and $\al$ is different than one leading to a different behaviour of the universe with the presence of the moduli field. This can also be observed by the discrepancy between $H$ and $H_m$ in fig.3c. However, it is important to remember that in this case  $V'|=0$ but $V''| \raw \infty$ giving an infinite mass for the moduli at $<T>=2$. 

\begin{figure}
\psfig{file=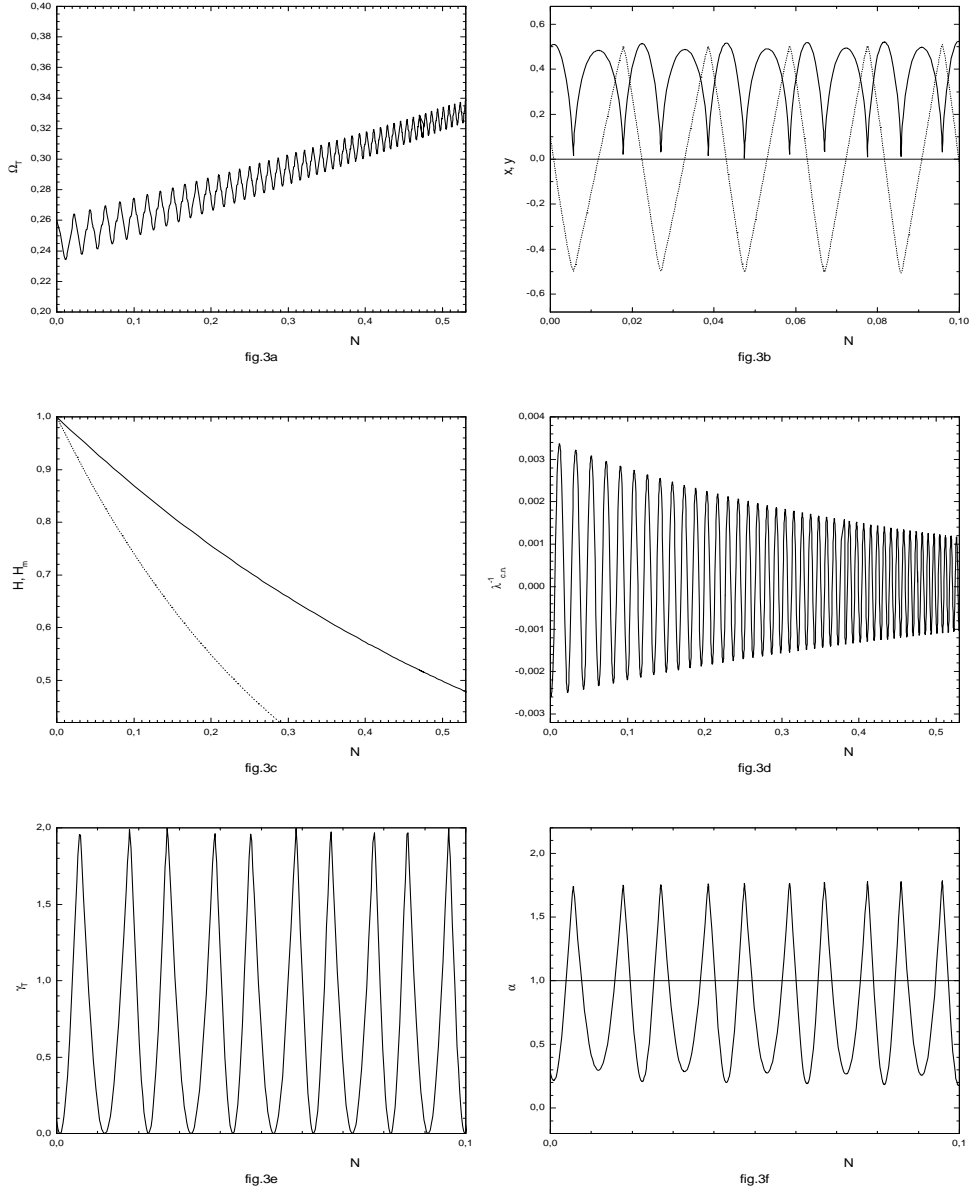,width=16cm,height=20cm}
\caption{
\footnotesize{
We show the evolution of different cosmological parameters for a potential $V=c\, T_r^2 |j'|^2|(j-j_0)|^{-7/15}$ with $j_0=j(2)$ and $T \raw 2$ at late times. The initial conditions are $x_0=0.1, y_0=0.5$ and $H_0=1$. In fig.3a we show the evolution of $\Omega_T$. In fig.3b we see the behaviour of x (dot line) and y (solid line). In fig.3c we plot the Hubble parameter (solid line) compared to a matter dominated case $H_m$  (dot line) while in fig.2d we have $\lm{c.n.}$. Finally we show in fig.1e and 1f the  $\gm_T$ parameter and the acceleration parameter $\al$ respectively.}}
\end{figure}

To conclude, we have studied the cosmological evolution of string moduli fields in a completely general way. We have only assumed the well established results of T-duality and the sigma model kinetic term for the moduli field (at large $T$). The scalar potential is model dependent but duality invariance constraints the potential severely. If the superpotential is given as a finite expansion on $j$ then, in the region of large $T$, the scalar potential has a double exponential form  leading to an  energy density redshift of the modular field  faster than the barotropic fluid. This result excludes this kind of potentials as a slow varying cosmological constant. Furthermore, for a finite v.e.v. of the moduli,  if its mass  is finite (i.e. $m(T_0) < \infty$), then the redshift of the moduli will be no slower than matter fields and will not be able to contribute to solve the age crises nor the positively accelerating universe. However, if we allow the moduli mass to be infinite then the redshift will be slower than for matter fields and it will give the main contribution to the energy density of the universe but it will not lead to a positively accelerating universe. We have therefore shown  that moduli field {\it can not} parameterize a slow rolling cosmological constant with a positively accelerating universe.

I am greatful  to G. Piccinelli for many useful discussions. This research was supported in part by CONACYT project 32415-E  and by DGAPA, UNAM,  project IN-103997.

\vskip 2truecm

{\bf Appendix}

\footnotesize{
In this appendix we give some useful relations between modular functions \ci{modfunc}. A modular function $f$ with weight $m$ transforms under duality $SL(2,Z)$ as  $f(T) \raw (icT+d)^m f(T)$ for $T\raw (aT-ib)/(icT+d),\,ad-bc=1,\, a,b,c,d \,\epsilon \,Z$. This symmetry has two dual invariant points $T=1, \delta=e^{i\pi/6}$ plus $SL(2,Z)$ related points. A complete set of independent modular (holomorphic) functions is given by the Eisenstein series, with modular weight $2k$,
\bea
G_{2k}(T) &=& \Sigma_{n_1,n_2 \epsilon Z}(i n_1 T + n_2)^{-2k}\non \\
G_{2k}(T)&=& 2\zeta (2k) \le( 1-\fr{4k}{B_{2k}}\Sigma_{n=1}^{\infty} \sigma_{2k-1}(n)q^n\ri)
\la{[i]}
\eea
where $q=e^{-2\pi T}$, $\zeta$ is the Riemann's zeta function, $B_{2k}$ are the Bernoulli numbers and $\sigma_p(n)$ is the p-powers of all divisors of n.
The Eisenstein functions (for $k >1$)  do not have any poles and their zeros are at $T=1$ for k odd and at $T=\delta$ for $k\neq 0 \,mod\, 3$. For $k=1$
the order of summation must be specified and one gets either $G_2(T)$, which is holomorphic but transforms as a connection $G_2 \raw (icT+d)^2 G_2-2\pi ic(icT+d)$, or   $\hat G_2(T,\bar T)=G_2(T)-\pi/Re T $, which transforms as a modular function with weight two but it is not holomorphic. $\hat G_2$ does not have any poles and its zeros are at the dual invariant points $T=1,\,\delta=e^{i\pi/6}$.

The derivatives of $G_{2k}$ for $k>1$ and $k=1$ are given by
\bea
G_{2k}'(T)&=& a(k) G_{2k+2}-\fr{k}{\pi}G_2G_{2k} \non\\
G_{2}'(T)&=& \fr{5}{2\pi} G_{4}-\fr{1}{2\pi}G_2^2
\la{a1}\eea
with $a(k)=\pi k\zeta(2k)/3\zeta(2k+2)$.
The derivative of any modular invariant function vanishes at the self dual points.
We can express the Dedekind Eta function $\eta(T)$, the absolute invariant function $j(T)$ and their derivatives  in terms of the Eisenstein functions 
\bea
\eta(T)=  \Delta^{1/24}=q^{1/24}\Pi_n(1-q^n), & \hspace{2cm}& \eta'(T)= -\fr{1}{4\pi}\eta \,G_2   \\
j(T)= \fr{3^65^3}{\pi^{12}}\fr{G_4^3}{\Delta},&\hspace{2cm }  & j'(T)= \fr{21}{\pi} j \fr{G_6}{G_4}
\la{a2}
\eea
with the discriminant
\be
\Delta(T)\equiv \fr{675}{256 \, \pi^{12}}\le( 20G_4^3-49G_6^2 \ri).
\ee
 The $\eta$ function has a zero at $T=\infty$ and its dual invariant point $T=0$ and it has no poles. The invariant function $j(T)$ has a triple zero at $T=\delta$ and it has a pole at $T=\infty$. Using eqs.(\ref{a1}) and (\ref{a2}) we can get some useful relationships 
\bea
G_4(T) = \frac{\pi^4}{45} \eta^{8} j^{1/3}, &\hspace{2cm} & G_4'(T)= \fr{7}{\pi}G_6 -\fr{2}{\pi}G_2 G_4 \non \\
G_6(T) =\frac{2 \pi^{6} \eta^{12}}{945} (j - 1728)^{1/2}, &\hspace{2cm}& G_6'(T)=\fr{10}{\pi}G_8 -\fr{3}{\pi} G_2 G_6 
\la{[iv]}
\eea
For finite $T$, $G_4$ has its only zero at $T=\delta$ while $G_6$  at $T=1$ and they do not have any poles.  Using the properties of the invariant function $j$ and of the Eisenstein function $G_{2k}$  one finds 
\be
\fr{j''}{j'}= \fr{21}{\pi}\fr{G_6}{G_4}+\fr{10}{\pi}\fr{G_8}{G_6}-\fr{7}{\pi}\fr{G_6}{G_4}-\fr{1}{\pi}G_2.
\la{jj/j}
\ee
The expansion around $T \gg 1$ gives
\bea
G_2 (T)&=& \frac{\pi^2}{3} (1-24 q - 72 q^2 - ...) \non \\
G_4 (T)&=& \frac{\pi^4}{45} (1+240 q + 2160 q^2 + ...) \non\\
G_6(T) &=& \frac{2 \pi^6}{945} (1-504 q - 16632 q^2 - ...) \\
j(T) &=& \frac{1}{q} + 744 + 196884 q + ... \non \\
\eta(T) &=& q^{1/24} (1 + q + 2 q^2 + 3q^3+  ...)^{-1} \non 
\la{[iii]}
\eea
To summerize, the values of $G_2,G_4,G_6,\eta, j, j', j''$ at $T=1,\delta,\infty$ are
\begin{center}
\begin{tabular}{|c|c|c|c|c|c|c|c|}
\hline 
  T & $G_2(T)$ & $G_4(T)$ & $G_6(T)$ & $\eta(T)$ & $ j(T)$ & $ j'(T)$ & $ j''(T)$ \\
\hline
1 & $\pi$ & $3.15\, \pi^4$ & 0 & $0.768\, \sqrt{2\pi}$ & 1728  &  0 & 49655\\
\hline
$\delta=e^{i\pi/6}$ & $2\pi/\sqrt{3}$  & 0  & $5.86\, \pi^6$ & $(0.79-i\, 0.10)\sqrt{2\pi} $ & 0 (triple) & 0 (double) & 0 \\
\hline
$\infty$ & $\pi^2/3$ & $\pi^4/45$ & $ 2\pi^6/945$ & $q^{1/24}\raw 0$ & $1/q \raw \infty$ & $ 2\pi/q \raw \infty$ & $4\pi^2/q\raw \infty$ \\
\hline 
\end{tabular}
\end{center}
}

\thebibliography{}

\footnotesize{
\bibitem{Riess} {A.G. Riess {\it et al.}, Astron. J. 116 (1998) 1009; S.
Perlmutter {\it et al}, ApJ 517 (1999) 565; P.M. Garnavich {\it et al}, Ap.J 509 (1998)74.}
\bibitem{Efst} {G. Efstathiou, S.L. Bridle, A.N. Lasenby, M.P. Hobson and R.S. Ellis,MNRAS 303L (1999) 47; M. Roos and S.M. Harun-or Rashid, astro-ph/9901234.}
\bibitem{Blinn} {E.I Sorokina, S.I. Blinnikov and O.S. Bartunov (astro-ph/9906494); A.G. Riess, A.V. Filippenko, W. Li and B.P. Schmidt, 
astro-ph/9907038.}
\bibitem{White} S.D.M. White, J.F. Navarro, A.E. Evrard and C.S. Frenk, Nature 366 (1993) 429.
\bibitem{Steig} G. Steigman and J.E. Felten, {\it Proceedings of the St. Petersburg Gamow Seminar}, ed. A.M Bykov and R.A. Chevalier, Sp. Sci. Rev. (1995)
\bibitem{EMS} G. Efstathiou, S. Maddox and W. Sutherland, Nature 348 (1990) 705.
\bibitem{klypin} {J. Primack and A. Klypin, Nucl. Phys. Proc. Suppl. 51 B, (1996), 30; J.F. Navarro and M. Steinmetz ,  astro-ph/9908119.}
\bibitem{Freed} W.L. Freedman, {\it David Schramm Memorial Volume}, Phys. Rep., in press (2000),  astro-ph/9909076.
\bibitem{Ferr} {L. Ferrarese {\it et al.}, {\it Proceedings of the Cosmic Flows Workshop} (1999), astro-ph/9909134 ; M.A. Hendry and S. Rauzy, {\it Proceedings of the Cosmic Flows Workshop} (1999),  astro-ph/9908343.}
\bibitem{Chab0} B. Chaboyer, P. Demarque, P.J. Kernan and L.M. Krauss, Science 271 (1996) 957.
\bibitem{Chab} B. Chaboyer, P. Demarque, P.J. Kernan and L.M. Krauss, ApJ 494 (1998) 96.
\bibitem{RatP} {B. Ratra and P.J.E. Peebles, Phys. Rev. D37 (1988) 3406} 
\bib{gral}{J.M
Overduin and F.I. Cooperstock, Phys. Rev. D58 (1998) 20;  V. Mendez, Class Quant. Grav. 13(1996) 3229;  J. Uzan, Phys. Rev. D59, (1999) 123510; A.P. Billyard, A.A. Coley, R.J. van den Hoogen, J. Ibanez and I. Olasagaste, gr-qc/9907053}
\bib{scherr}{ A.R. Liddle and R.J. Scherrer,Phys. Rev. D59,  (1999) 023509}
\bib{mioscalar} A. de la Macorra and G. Piccinelli, hep-ph/9909459
\bib{cc+redsh}{W. Chen and Y. Wu, Phys. Rev D41 (1990) 695;  Y. Fujui and T. Nishioka, Phys. Rev. D42 (1990) 361; D. Pavon, Phys. Rev. D43 (1991) 375; J. Matyjasek, Phys. rev. D51 (1995) 4154; M. S. Berman,  Phys. Rev. D43 (1991) 1075; J.C. Carvalho, J.A.S. Lima and I. Waga, Phys. Rev. D46 (1992) 2404; J.A.S. Lima and J.M.F. Maia, Phys. Rev. D49 (1994) 5597; V. Silveira and I. Waga, Phys. Rev D50 (1994) 489} 
\bibitem{Olson} {T.S. Olson and T.F. Jordan, Phys. Rev. D 35 (1987) 3258; J. W. Moffat, Phys. Lett. B357 (1995) 526}
\bibitem{PeebR} {P.J.E. Peebles and B. Ratra, ApJ 325 (1988) L17; J.W. Moffat, Phys. Lett. B 357 (1995) 526.}
\bibitem{Freese} {K. Freese, F.C. Adams, J.A. Frieman and E. Mottola, Nucl. Phys. B 287 (1987) 797; M. Birkel and S. Sarkar, Astropart. Phys. 6 (1997) 197.} 
\bib{Wet}{C. Wetterich, Astron. Astrophys.301 (1995) 321
\bibitem{Cald} R.R. Caldwell, R. Dave and P.J. Steinhardt, Phys. Rev. Lett. 80 (1998) 1582.
\bibitem{Silv} {V. Silveira and I. Waga, Phys. Rev. D (1997) 4625; L. Amendola  astro-ph/9908023.}
\bib{quint}{D. Lyth and C. Koldo, Phys. Lett. B458 (1999) 197; I. Zlater, L. Wang and P.J. Steinhardt, Phys. Rev. Lett.82 (1999) 896; D. Huterer and M.S. Turner astro-ph/9808133; P. Brax and J. Martin, astro-ph/9905040; T. Chiba, gr-qc/9903094}
\bibitem{Turn} {M.S. Turner and M. White Phys. Rev. D 56 (1997) 4439; G. Efstathiou,  astro-ph/9904356; L. Wang {\it et al}, astro-ph/9901388.}
\bibitem{Albr} A. Albrecht and  C. Skordis, astro-ph/9908085. 
\bib{M}{ E. Witten, Nucl. Phys. B500 (1997)3; Nucl. Phys. B463 (1996) 383; J. Schwarz, Phys. Lett. B367 (1996)97; O. Aharony, J. Sonnenschein and S. Yankielowicz, Nucl. Phys. B474 (1996) 474}
\bibitem{Wet2} C. Wetterich, Nucl. Phys. B302 (1998) 668 
\bibitem{Vexp} {P. Ferreira, M. Joyce, Phys. Rev. D 58 (1998) 503; E.J. Copeland, A. Liddle and D. Wands, Ann. N.Y. Acad. Sci. 688 (1993) 647.   }
\bib{liddle}E.J. Copeland, A. Liddle and D. Wands, Phys. Rev. D57 (1998) 4686
\bib{moduli}{T. Banks, M. Berkooz and P. J. Steinhardt Phys. Rev. D52 (1995)705 }
\bib{Tdual}{J.J.Atick and E. Witten, Nucl. Phys B310 (1988)291; V.P. Nair, A. Shapere, A. Strominger and F. Wilczsk, Nucl. Phys. B287 (1987)402; A. Givon,
E. Rabinovici and G. Veneziono, Nucl. Phys. B322 (1989)167; A. Alvarez and M.A.R. Osorio, Phys. Rev. D40 (1989)1150}
\bib{sl2z}{A. Shapere and F. Wilczsk, Nucl. Phys. B320 (1989)669; L.E. Ibanez and D. Lust, Nucl. Phys. B382 (1992) 305} 
\bib{moduli/m1}{D. Bernstein, R. Corrado and J. Distler, Nucl. Phys.B503 (1997) 239, A. Lukas and B. Ovrut Phys. Lett. B437 (1998) 291}
\bib{moduli/m2}{P. J. Steinhardt hep-th/9907080; T. Banks hep-th/9906126}  A. Lukas, B.Ovrut and D. Waldram, Nucl. Phys. B509 (1998) 169}
 \bib{modfunc}{J. Lehner, Discontinous groups and automorphic functions. The American Mathematical Society (1964)}
\bib{miolog} A. de la Macorra, in preparation
}

\end{document}